\documentclass[a4paper]{scrartcl} 

\usepackage[english]{babel} 

\usepackage[sc,osf]{mathpazo}

\setkomafont{section}{\normalfont\bfseries\Large} 
\setkomafont{subsection}{\normalfont\bfseries\large} 
\setkomafont{subsubsection}{\normalfont\bfseries\normalsize} 
\setkomafont{paragraph}{\normalfont\bfseries\normalsize}
\setkomafont{title}{\normalfont} 

\usepackage[utf8]{inputenc} 
\usepackage[T1]{fontenc}

\usepackage{amsmath} 
\usepackage{amssymb}
\usepackage{amsthm}
\usepackage{stmaryrd}
\usepackage{url}
\usepackage{color}

\usepackage{tikz} 
\usetikzlibrary{trees,decorations,arrows,positioning,plotmarks,shapes}

\usepackage{enumerate}

\usepackage{listliketab}

\usepackage[bottom=4.5cm,inner=4cm,outer=3cm,marginparwidth = 2.1cm,
headsep = 1.5cm
]{geometry}

\usepackage{hyperref}
\hypersetup{
     colorlinks   = false,
     pdfauthor = Tim Jungnickel and Tobias Herb,
     pdftitle = TP1-valid Transformation Functions for Operations on ordered n-ary Trees
}

\newcounter{sdefinition}
\newenvironment{sdefinition}[2]{
\par\addvspace{6pt}
\refstepcounter{sdefinition}%
\label{#1}
\noindent\textbf{Definition~\arabic{sdefinition} (#2).}
\itshape
}{\par\addvspace{6pt}}

\newcounter{slemma}
\newenvironment{slemma}[1]{
\refstepcounter{slemma}%
\label{#1}%
\par\addvspace{6pt}\noindent\textbf{Lemma \arabic{slemma}.}
\itshape
}{ \par\addvspace{6pt}}

\newenvironment{lproof}[1]{
 \par\noindent\textbf{Proof of \hyperref[#1]{Lemma}~\ref{#1}.}
\parindent 0pt
}{ \par \vspace*{6pt}}

\newenvironment{tproof}[1]{
 \par\noindent\textbf{Proof of \hyperref[#1]{Theorem}~\ref{#1}.}
\parindent 0pt
}{ \par \vspace*{6pt}}

\newcounter{case}
\newenvironment{case}{
\par\addvspace{6pt}
\refstepcounter{case}
\textit{Case \arabic{case}.}}{}

\newcounter{subcase}[case]
\newenvironment{subcase}{
\par\addvspace{6pt}
\refstepcounter{subcase}
\textit{Case \arabic{case}.\arabic{subcase}.}}{}

\newcounter{stheorem}
\newenvironment{stheorem}[1]{
\refstepcounter{stheorem}%
\label{#1}%
\vspace{6pt}  \par\noindent\textbf{Theorem \arabic{stheorem}.}
\itshape
}{ \par \vspace*{6pt}}

\newcommand{\IL}{\operatorname{insert}_L}
\newcommand{\emptylist}{\left[ \,\right]}
\newcommand{\IT}{\operatorname{insert}_T}
\newcommand{\DT}{\operatorname{delete}_T}
\newcommand{\DL}{\operatorname{delete}_L}
\newcommand{\XFORM}{\operatorname{XFORM}}
\newcommand{\TPt}{\operatorname{TPt}}

\newcommand{\UP}{\operatorname{update}}
\newcommand{\noop}{\operatorname{no-op}}

\newcommand{\defref}[1]{\hyperref[#1]{Def.}~\ref{#1}}
\newcommand{\tabref}[1]{\hyperref[#1]{Table}~\ref{#1}}
\newcommand{\figref}[1]{\hyperref[#1]{Fig.}~\ref{#1}}
\newcommand{\lemref}[1]{\hyperref[#1]{Lemma}~\ref{#1}}

\newcommand{\lstref}[1]{\hyperref[#1]{Listing}~\ref{#1}}
\newcommand{\linenumber}[1]{\renewcommand{\ttdefault}{DejaVuSansMono-TLF}{\scriptsize \texttt{#1}}\renewcommand{\ttdefault}{lmtt}}

\definecolor{light-gray}{gray}{0.75}

\usepackage{listings}

\title{TP1-valid Transformation Functions for Operations on ordered $n$-ary Trees}
\author{Tim Jungnickel and Tobias Herb\\TU Berlin}

\begin{document} 

\date{}
\maketitle

\begin{abstract}
Collaborative work on shared documents was revolutionized by web services like Google Docs or Etherpad. Multiple users can work on the same document in a comfortable and distributed way. For the synchronization of the changes a replication system named Operational Transformation is used. Such a system consists of a control algorithm and a transformation function. In essence, a transformation function solves the conflicts that arise when multiple users change the document at the same time. In this work we investigate on the correctness of such transformation functions. We introduce transformation functions $n$-ary trees that we designed especially for the purpose of synchronization changes on JSON objects. We provide a detailed proof of the necessary property: the Transformation Property 1.
\end{abstract}

\section{Introduction}\label{sec:introduction}

The collaborative work on shared documents has become extraordinary comfortable by web services like Google Docs\footnote{\url{https://docs.google.com}} or Etherpad\footnote{\url{http://etherpad.org/}}. Multiple users can edit a shared document at the same time and all changes will be automatically synchronized in the background. As comfortable the usage of such services is, as challenging and non trivial is the synchronization process in the background. Users demand a consistent result i.e., no matter which user modifies the document, the result after the synchronization must identical for all users. An optimistic replication system named \emph{Operational Transformation} (OT) is used to guarantee that all replicas are consistent. Within this work extend our approach we present in \cite{JuHeACM} and provide the missing preliminaries that are necessary to build a fully working OT synchronization of JSON objects.

An OT system consists of two core components: a control algorithm and a transformation function \cite{Sun:2014:ESP:2531602.2531630}. A control algorithm handles the communication between the clients and the exchange of the changes. A transformation function solves the conflicts that occur when two clients independently edit a document at the same time.

To give some intuition of the OT mechanism, we demonstrate OT with a simple text editing scenario. The sites S1, S2 replicate the character sequence \texttt{XYZ}. S1 inserts character \texttt{A} at position 0, resulting in \texttt{AXYZ}. S2 simultaneously deletes the character \texttt{Y} at position 1, resulting in \texttt{XZ}. The operations are exchanged between the sites. If the remote operations are applied naively, then we get inconsistent replicas: S1 results in \texttt{AYZ} and S2 results in \texttt{AXZ}. Therefore, a transformation of the remote operation is performed, before the operation is applied. At S1, the position of the remote delete operation is incremented with respect to the local insert operation. At S2, the remote insert operation does not need to be transformed since the local delete operation has no effect on the remote insert operation. The transformations of the position of simultaneous operations leads to consistent replicas.

The major problems of applying linear OT on hierarchical structures are simultaneous inserts of hierarchical nodes. This can be well exemplified by an HTML example. We reuse our document containing the sequence \texttt{XYZ}, replicated on the sites S1, S2. Site S1 decides that character \texttt{Y} should be in bold. This can be represented with two insert operations, producing the document state \texttt{X<b>Y</b>Z}. Simultaneously S2 decides that character \texttt{Y} should be in italic. If the server receives S2's edit after S1, it would need to transform the operations of S2 against the operations of S1. Usually the server would be configured to place the later edit behind the first edit. If the operations of S1 are applied before the transformed operations of S2 are applied, then the document will be syntactically incorrect: \texttt{X<b><i>Y</b></i>Z}.

\paragraph{Contributions}

The contributions of this work are the following:
\begin{itemize}
\item We verify the necessary properties of a common transformation function for operations on lists with a Isabelle/HOL proof.
\item We introduce a transformation function for operations on $n$-ary trees that is designed to support simultaneous editing of JSON objects in a programming language near notation.
\item We verify the necessary properties of the introduced transformation function with a detailed proof.
\end{itemize}

\section{Related Work}

Davis et al~\cite{Davis:2002:GOT:587078.587088} were, to our knowledge, the first that applied the OT approach on treelike structures. They extended operational transformation to support synchronous collaborative editing of documents written in dialects of SGML (Standard General Markup Language) such as XML and HTML. They introduced a set of structural operations with their associated transformation functions tailored for SGML's abstract data model grove. Their approach is followed by \cite{4530334,Sun:2006:TAS:1188816.1188821} showing improvements in XML editing and implementations in collaborative business software. In our work we provide an alternative, more generic transformation function that is not shaped for XML. Hence with our transformation function we enable more general use cases of hierarchical OT. From this we can easily derive a transformation of operations on JSON objects. In addition we present our transformation functions in a programing language near notation so that they can be implemented easier.

Oster et al~\cite{OsterICEIS07} proposed a framework for supporting collaborative writing of XML documents. Their framework works similarly to the Copy-Modify-Merge paradigm widely used in version control systems such as CVS. The synchronization of the replicated XML documents is based on Operational Transformation. They make also use of positional addressing scheme of the XML elements, comparable to our approach. They claim with respect to proving the correctness: ``It is nearly impossible to do this by hand'' and refer to an automated tool named VOTE to fulfill the challenge~\cite{Imine2006167}. Compared to their work we define the underlying model precisely for our purpose --- the synchronization of changes on generic hierarchical objects. Moreover we took the challenge of Oster et al and managed to prove the correctness \emph{by hand}.

Our approach is most comparable with the approach of Imine et al~\cite{Imine:2003:PCT:1241889.1241904}. They analyzed the correctness of common transformation functions for lists with respect to the necessary transformation properties with SPIKE. Based on their observations they propose a new transformation function with the desired properties that is easier to read and to prove. We also took one common transformation function for lists and used the interactive theorem prover Isabelle/HOL to verify one transformation property. Based on the gained experience we proposed a new transformation function for operations on $n$-ary trees that is, in contrast to the described work in the previous paragraphs, easier to understand, prove and implement.

\section{Technical Preliminaries}

In this section we present the necessary preliminaries to define a transformation function for tree operations.

\subsection{Lists}

We consider lists as one of the simplest form of a linear data structure. A list is recursively defined either as an \emph{empty list} or as a \emph{cell} containing an item and another list. In \tabref{tab:notation} we introduce a custom notation for lists similar to the notation in the programming language Python. Based on this notation we define an insert and a delete operation for lists in \defref{def:insertL} and \defref{def:deleteL}.

\begin{table}[t]
\caption{Notation for lists based on the programming language Python.}\label{tab:notation}
\centering
\resizebox{\columnwidth}{!}{%
\begin{tabular}{p{1.8cm} p{12cm}}
\hline\noalign{\smallskip}
Notation & Description\\
\noalign{\smallskip}
\hline 
\noalign{\smallskip}
\centering $\emptylist$ & \textbf{Empty List and Delimiters $\;$} We use $[$ and $]$ as delimiters for a list. Hence the list $\left[x,y,z\right]$ contains the items $x$, $y$ and $z$. We denote the empty list as $\emptylist$. \\
\centering$|L|$ & \textbf{Length of a List $\;$} We define the length of a list $L$ as the number of items in the list, denoted as $|L|$. \\
\centering$L[n]$ & \textbf{List Access $\;$} Let $L$ be an arbitrary list. We denote the access to the $n$\textsuperscript{th} element as $L[n]$. Note that $L[n]$ is only defined if $n < |L|$. We access the first element with $L[0]$. \\
\centering$L[x,y]$ & \textbf{Intervals $\;$} Let $L$ be an arbitrary list. We write $L[x,y]$ for a new list that contains all elements from $L[x]$ to $L[y]$. Note that $L[x,y]$ can be the empty list if $x > y$ or the if $x$ and $y$ referencing to non existing elements. If only $y$ is referencing to a non existing element, $L[x,y]$ contains all elements from $L[x]$ to the last element of $L$. \\
\centering$L_1 + L_2$ & \textbf{List Concatenation $\;$} Let $L_1$ and $L_2$ be two arbitrary lists. We define the concatenation of $L_1$ and $L_2$, denoted as $L_1 + L_2$, as a new list that starts with $L_1$ and ends with $L_2$. \\
\centering$L_1 \subseteq L_2$ $L_1 \subset L_2$& \textbf{Sublists and Strict Sublists $\;$} Let $L_1$ and $L_2$ be two arbitrary lists. We call $L_1$ a \emph{sublist} of $L_2$, denoted as $L_1 \subseteq L_2$, if $L_2$ starts with $L_1$. If $L_1 \subseteq L_2$ and $|L_1| < |L_2|$, we call $L_1$ a \emph{strict sublist} of $L_2$, denoted as $L_1 \subset L_2$. \\
\centering$L[\le x]$ $L[< x]$ & \textbf{Head and Tail $\;$} Let $L$ be an arbitrary list. We write $L[\le x]$ for a new (sub)list that contains all elements from the first element to $L[x]$. We use the abbreviation $L[< x]$ for $L[\le x - 1]$. The lists $L[\ge x]$ and $L[> x]$ are defined analogously.\\
\hline
\end{tabular}}
\end{table}

\begin{sdefinition}{def:insertL}{$\IL$}
The operation $\IL$ has three parameters: an item $i$, a position $k$ and a list $L$ with $k \le |L|$. As result, the item $i$ is inserted into the list $L$ at position $k$: \[\IL(i,k,L) \triangleq L[ < k ] + [i] + L[\ge k] \]
\end{sdefinition}

\begin{sdefinition}{def:deleteL}{$\DL$}
The operation $\DL$ has two parameters: a position $k$ and a list $L$ with $k < |L|$. As result, the item at position $k$ is deleted from $L$: \[\DL(k,L) \triangleq L[ < k ] + L[> k] \]
\end{sdefinition}

\subsection{Trees}\label{sec:trees}
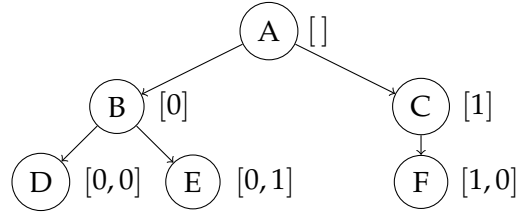
\begin{figure}[ht]
  \centering
    \begin{tikzpicture}[auto,bend angle=30,scale=0.5]
      \node [draw, circle, minimum height=0.6cm](A) at (0,0)  {A};
      \node [draw, circle, minimum height=0.6cm](C) at (4,-2) {C};
      \node [draw, circle, minimum height=0.6cm](B) at (-4,-2) {B};
      \node [draw, circle, minimum height=0.6cm](D) at (-6,-4) {D};
      \node [draw, circle, minimum height=0.6cm](E) at (-2,-4) {E};
      \node [draw, circle, minimum height=0.6cm](F) at (4,-4) {F};
    
      \node at (1.3,0) {$\emptylist$};
      \node at (-2.5,-2) {$\left[ 0 \right]$};
      \node at (5.5,-2) {$\left[ 1 \right]$};
      \node at (-4.1,-4) {$\left[ 0,0 \right]$};
      \node at (-0.1,-4) {$\left[ 0,1 \right]$};
      \node at (5.8,-4) {$\left[ 1,0 \right]$};
      \path[->] (A) edge node {} (B);
      \path[->] (A) edge node {} (C);
      \path[->] (B) edge node {} (D);
      \path[->] (B) edge node {} (E);
      \path[->] (C) edge node {} (F);

    \end{tikzpicture}
  \caption{Tree representation and node access}
  \label{fig:tree-access}
\end{figure}

We consider $n$-ary trees with the simplest set of operations $\IT$ and $\DT$. An $n$-ary tree is recursively defined as a pair of a value and a list of trees. A leaf is defined as a pair of a value and an empty list. Thus a tree cannot be empty, the smallest tree is a single leaf. As shown in \figref{fig:tree-access}, we use a list of natural numbers (called \emph{access path}) to access the tree at a specific position. For a tree $T$ and an access path $pos$ we write $T\llbracket pos \rrbracket$ to get the \emph{subtree} at position $pos$. We define the operations $\IT$ and $\DT$ in \defref{def:insertT} and \defref{def:deleteT}.

\begin{sdefinition}{def:insertT}{$\IT$} 
The operation $\IT$ has three input parameters: a tree $t$, a non empty access path $pos$ and a tree $T = (v,L)$. As result, the tree $t$ will be inserted into $T$ at position $pos$. We define the operation recursively:

\begin{tabular}{l}
$\IT(t,[x],(v,L)) \triangleq (v,\IL(t,x,L))$ \vspace*{0.1cm} \\ 
$\IT(t,[x] + xs,(v,L)) \triangleq (v,\IL(\IT(t,xs,L[x]),x,\DL(x,L)))$
\end{tabular}
\end{sdefinition}

\begin{sdefinition}{def:deleteT}{$\DT$}
The operation $\DT$ has two input parameters: a non empty access path $pos$ and a tree $T = (v,L)$. As result, the subtree at position $pos$ will be deleted from $T$. We define the operation recursively:

\begin{tabular}{l}
$\DT(t,[x],(v,L)) \triangleq (v,\DL(x,L))$ \vspace*{0.1cm} \\ 
$\DT(t,[x] + xs,(v,L)) \triangleq (v,\IL(\DT(xs,L[x]),x,\DL(x,L)))$
\end{tabular}
\end{sdefinition}

According to our definition of trees, the second last element of an access path determines the node where a subtree should be inserted into or where a subtree should be deleted from. The last element of an access path determines the position inside the list of subtrees of the node at the second last element. We simply use the operation $\IL$ to insert a tree into the list of subtrees and we use $\DL$ to delete a tree from the list of subtrees.

We notice that both definitions of $\IT$ and $\DT$ are insufficient if the input access path directs to a non existing node. For the rest of the paper we assume, that we have a \emph{valid} input for both operations i.e., the access path directs to a position where we can apply an insertion or a deletion. For a later setup in a server/client architecture we can easily check on both sides if an operation is valid. We define the validity of tree operations more precisely in \defref{def:validT}.

\begin{sdefinition}{def:validT}{Valid Tree Operations}
  Let $O$ be a tree operation on the tree $T$. We call $O$ a \emph{valid tree operation} if:
  \begin{itemize}
    \item \textbf{case 1:} $O = \IT(t,pos,T)$, then:
    
    Up to the second last element the position parameter $pos$ directs to an existing subtree $(v,L)$ in $T$ and the last element of $pos$ is a valid position parameter for an $\IL$ operation in $L$.
    \item \textbf{case 2:} $O = \DT(pos,T)$, then:
    
    The position parameter $pos$ directs to an existing subtree in $T$.
  \end{itemize}
\end{sdefinition}

An $\IT$ operation can be considered for any strict sublist of the access path. As demonstrated in \figref{fig:Lemma1}, the original $\IT$ operation on the left of the equation can also be seen as an $\IT$ operation with a sublist of the original access path, as long as the emerging subtree (visualized as thick circles and arrows) and the deleted subtree (visualized as dotted circles and arrows) are adjusted. We formalize this observation in \lemref{lem:insertT}. \lemref{lem:deleteT} is the counterpart for $\DT$ to \lemref{lem:insertT}. 

\begin{figure*}[ht]
  \centering
    \begin{tikzpicture}[auto,bend angle=30,node distance=2cm,scale=0.7]
      \node [draw, circle, minimum height=.6cm](A) at (0,0) {A};
      \node [draw, circle, minimum height=.6cm](C) at (-1,-1.5) {B};
      \node [draw, circle, minimum height=.6cm](B) at (1,-1.5) {C};
      \node [draw, circle, minimum height=.6cm](D) at (1,-3) {D};
      \node [draw, circle, minimum height=.6cm](E) at (-4.4,-2) {E};
      \node (IT1) at (-6.1,-2) {$\IT \big ($};
      \node (pos) at (-2.7,-2) {$, \left[1,0,0\right] ,$};
      \node at (2,-2) {$\big ) $};
      \node at (2.7,-2) {$=$};
      
      \path[->] (A) edge node {} (B);
      \path[->] (A) edge node {} (C);
      \path[->] (B) edge node {} (D);
      
      \node [draw, circle, minimum height=.6cm](A2) at (9.5,0)  {A};
      \node [draw, circle, minimum height=.6cm](C2) at (8.5,-1.5) {B};
      \node [draw, circle, minimum height=.6cm,dotted](B2) at (10.5,-1.5) {C};
      \node [draw, circle, minimum height=.6cm,dotted](D2) at (10.5,-3) {D};
      \node [draw, circle, minimum height=.6cm,thick](C3) at (6.1,-0.5) {C};
      \node [draw, circle, minimum height=.6cm,thick](D3) at (6.1,-2) {D};
      \node [draw, circle, minimum height=.6cm,thick](E3) at (6.1,-3.5) {E};
      \node  at (4.4,-2) {$\IT \big ($};
      \node  at (7.3,-2) {$, \left[1\right] ,$};
      \node at (11.5,-2) {$\big ) $};
      
      \path[->,dotted] (A2) edge node {} (B2);
      \path[->] (A2) edge node {} (C2);
      \path[->,dotted] (B2) edge node {} (D2);
      \path[->,thick] (C3) edge node {} (D3);
      \path[->,thick] (D3) edge node {} (E3);
    \end{tikzpicture}
  \caption{Example for \lemref{lem:insertT}}
  \label{fig:Lemma1}
\end{figure*}

\begin{slemma}{lem:insertT}
Given a tree $T$ and a valid access path $pos$ for an $\IT$ operation. For all $0 < i < |pos|$ and for all $t$ the following statement is always true:

\begin{tabular}{l}
$\IT(t,pos,T) =$ \\
\hspace*{0.5cm}$\IT(\IT(t,pos\left[\ge i\right],T\llbracket pos\left[< i\right] \rrbracket), pos\left[< i\right],\DT(pos\left[< i\right],T))$
\end{tabular}
\end{slemma}

\begin{lproof}{lem:insertT}
  We prove \lemref{lem:insertT} by induction over $i$. 
  
\textbf{base case:} We fix an access path $pos$ and a tree $t$ and show the base case for $i = 1$ with $i < |pos|$ directly with the presented definitions. We assume $pos$ is a valid access path for an $\IT$ operation on $T =(v,L)$.
{\scriptsize 
\begin{eqnarray*}
&& \IT(t,pos,T)\\ [-1mm]
&\stackrel{\text{D}\ref{def:insertT}}{=} & (v,\IL(\IT(t,pos[\ge 1],L\bigl[pos[0]\bigr])),pos[0],\DL(pos[0],L))\\ [-1mm]
&\stackrel{\text{D}T}{=} & (v,\IL(\IT(t,pos\left[\ge 1\right],T\llbracket pos\left[< 1\right] \rrbracket)),pos\left[0\right],\DL(pos[0],L))\\ [-1mm]
&\stackrel{ \text{D}\ref{def:insertT}}{=} & \IT(\IT(t,pos\left[\ge 1\right],T\llbracket pos\left[< 1\right] \rrbracket),pos\left[< 1\right],(v,\DL(pos[0],L)))\\ [-1mm]
&\stackrel{ \text{D}\ref{def:deleteT}}{=} & \IT(\IT(t,pos\left[\ge 1\right],T\llbracket pos\left[< 1\right] \rrbracket),pos\left[< 1\right],\DT(pos\left[< 1\right],T))
\end{eqnarray*}} 
\textbf{inductive step:} We fix an access path $pos$ and a tree $t$ and show the inductive step with $(i+1) < |pos|$ directly with the presented definitions and the induction hypothesis. We assume $pos$ is a valid access path for an $\IT$ operation on $T$. Let $T \llbracket pos[< i]\rrbracket = (v,L)$.
{\scriptsize 
\begin{eqnarray*}
&& \IT(\IT(t,pos[> i],T \llbracket pos[\le i]\rrbracket),pos[\le i],\DT(pos[\le i],T))\\[-1mm]
&\stackrel{\text{IH}}{=} & \IT\bigl(\IT(\IT(t,pos[> i],T \llbracket pos[\le i]\rrbracket),(pos[\le i])[\ge i],\\[-1mm]
&& \hspace{1.5cm} \DT(pos[\le i],T)\llbracket pos[< i] \rrbracket),\\[-1mm]
&& \hspace{0.7cm}(pos[\le i])[< i],\DT(pos[< i],\DT(pos[\le i],T))\bigr)\\[-1mm]
&\stackrel{\text{D}\ref{def:deleteT}}{=} & \IT\bigl(\IT(\IT(t,pos[> i],T \llbracket pos[\le i]\rrbracket),[pos[i]],\\[-1mm]
&& \hspace{1.5cm} \DT(pos[\le i],T)\llbracket pos[< i] \rrbracket),\\[-1mm]
&& \hspace{0.7cm}pos[< i], \DT(pos[< i],T)\bigr)\\[-1mm]
&\stackrel{\text{D}T,\text{D}\ref{def:deleteT}}{=} & \IT\bigl(\IT(\IT(t,pos[> i],T \llbracket pos[\le i]\rrbracket),[pos[i]],\\[-1mm]
&& \hspace{1.5cm} (v,\DL(pos[i],L))),\\[-1mm]
&& \hspace{0.7cm}pos[< i], \DT(pos[< i],T)\bigr)\\[-1mm]
&\stackrel{\text{D}T,\text{D}\ref{def:insertT}}{=} & \IT\bigl((v,\IL(\IT(t,pos[> i],L\bigl[pos[i]\bigr]),pos[i],\\[-1mm]
&& \hspace{1.75cm} \DL(pos[i],L))),\\[-1mm]
&& \hspace{0.7cm} pos[< i], \DT(pos[< i],T)\bigr)\\[-1mm]
&\stackrel{\text{D}\ref{def:insertT}}{=} & \IT\bigl(\IT(t,pos[\ge i],(v,L)),pos[< i], \DT(pos[< i],T)\bigr)\\[-1mm]
&\stackrel{\text{IH}}{=} & \IT(t,pos,T)
\end{eqnarray*}} \qed
\end{lproof}

\begin{slemma}{lem:deleteT}
Given a tree $T$ and a valid access path $pos$ for a $\DT$ operation. For all $0 < i < |pos|$ the following statement is always true:

\begin{tabular}{l}
$\DT(pos,T) =$ \\
\hspace*{0.5cm}$\IT(\DT(pos\left[\ge i\right],T\llbracket pos\left[< i\right] \rrbracket), pos\left[< i\right],\DT(pos\left[< i\right],T))$
\end{tabular}
\end{slemma}

\begin{lproof}{lem:deleteT}
We can easily prove \lemref{lem:deleteT} with induction over $i$ similar to the proof of \lemref{lem:insertT}. \qed
\end{lproof}

In \lemref{lem:Telimination} we state, that an $\IT$ operation is eliminated by a $\DT$ operation if both operations have the same access path and that if we access a tree at a position where we just inserted an element, we receive exactly that element.

\begin{slemma}{lem:Telimination}
For a given tree $T$ and a valid access path $pos$ for an $\IT$ operation we know that:
\begin{enumerate}
  \item $\forall t . \DT(pos,\IT(t,pos,T)) = T$ \label{enum:deliteliminate}
  \item $\forall t . \IT(t,pos,T)\llbracket pos \rrbracket = t$ \label{enum:itfilter}
\end{enumerate}
\end{slemma}

\begin{lproof}{lem:Telimination}
The claims follow directly from the definitions of $\IT$ and $\DT$. \qed
\end{lproof}

\subsection{Operational Transformation and Transformation Functions}

We consider an \emph{operation} to be an identifier together with zero or more input values and well defined semantics. For example the $\IL$ operation has three input values: an item to be inserted, a position and a list. The semantics is given by the implementation. Sometimes we consider instances of an operation i.e., operations with given input values. For example $O_1 = \IL(a,2,[x,y,z])$. In this case we call the list $[x,y,z]$ the \emph{context} of this instance since the result would be a new context for another instance. We denote the context of one instance $O_1$ as $C(O_1)$. For the rest of this work we omit the term \emph{instance} and just say operation, even if concrete values are present.

\begin{sdefinition}{def:operationComposition}{Operation Composition}
Given two operations $O_1$ and $O_2$ where the context of $O_2$ is the result of $O_1$. We write $O_2 \circ O_1$ for the composition of the operations where $O_2$ is applied \emph{after} $O_1$ has been applied.
\end{sdefinition}

We write $O_2 \circ O_1 (X)$ to express that the context of $O_1$ is $X$.

\begin{sdefinition}{def:XFORM}{Transformation Function} A \emph{Transformation Function} has a pair of operation instances $(O_1,O_2)$ with $C(O_1) = C(O_2)$ as input parameter and returns a pair of transformed operation instances $(O_1',O_2')$ where $O_1'$ is the transformed version of $O_1$ with $C(O_1') = O_2$ and $O_2'$ is the transformed version of $O_2$ with $C(O_2') = O_1$.
\end{sdefinition}

Our definition of the transformation function is based on the introduction of the Jupiter OT system~\cite{Nichols:1995:HLW:215585.215706}. The input of a transformation function can be seen as two independent operations (more concrete: instances) of two sites. Both sites apply the operation to a local replica of a context which may result in inconsistent replicas over both sites. In order to achieve consistent replicas, the transformed versions of the operations are exchanged and applied by both sites.

In \lstref{lst:XFORML} we present one transformation function for list operations $\XFORM_L$ which was initially introduced by Ellis and Gibbs in \cite{Ellis:1989:CCG:66926.66963} and slightly improved by Ressel et al in \cite{Ressel:1996:ITA:240080.240305}. In the listing we omit the last parameter of each operation since the context of all operations is defined in \defref{def:XFORM}. We implemented a proof in the interactive theorem prover Isabelle to verify that the transformation is correct and that the replicas will be consistent eventually\footnote{\url{https://gitlab.tubit.tu-berlin.de/jungnickel/isabelle}}. 

The transformation of the "XYZ" example from the introduction would be processed in the lines \linenumber{7} and \linenumber{13} of \lstref{lst:XFORML}. According to line \linenumber{4} we need to use application dependent priorities to transform two $\IL$ operations with identical position parameters. For example we can define a priority where operations of one site are prioritized over all other operations. If we have two $\DL$ operations with identical position parameters (see line \linenumber{19}), both sites independently delete the same element from the list which result in a consistent state. Hence we transform both operations to $\noop$.

One essential property of the transformation function to achieve consistent replicas is the \emph{Transformation Property 1} (TP1) \cite{Ressel:1996:ITA:240080.240305}. In essence TP1 describes that the transformation function needs to repair the inconsistencies that occur if two operation instances are applied in different orders.

\begin{sdefinition}{def:TP1}{Transformation Property 1}
Let $O_1$ and $O_2$ be two arbitrary operations with the same context $C(O_1) = C(O_2)=\mathcal{C}$. A transformation function $\operatorname{XFORM}$ satisfies the \emph{Transformation Property 1} (TP1), if the following holds for $\operatorname{XFORM}(O_1,O_2)=(O_1',O_2')$: \[O_2' \circ O_1 (\mathcal{C}) = O_1' \circ O_2 (\mathcal{C})\] 
\end{sdefinition}

\renewcommand{\ttdefault}{DejaVuSansMono-TLF}

\begin{figure}[ht]
\lstset{
basicstyle=\scriptsize\ttfamily, 
frame=top bottom,
numbers=left,
numberstyle=\tiny\ttfamily,
language=Python,
commentstyle=\color{light-gray},
caption={Pseudo code of the transformation function $\XFORM_L$},
label={lst:XFORML},
floatplacement=t
}
\begin{lstlisting}
function XFORML(insertL(i1, k1), insertL(i2, k2)):  
  if k1 < k2: return(insertL(i1, k1), insertL(i2, k2 + 1))
  if k1 > k2: return(insertL(i1, k1 + 1), insertL(i2, k2))
  if k1 == k2: # use application dependent priorities
  
function XFORML(insertL(i, k1), deleteL(k2)):  
  if k1 < k2: return(insertL(i, k1), deleteL(k2 + 1))
  if k1 > k2: return(insertL(i, k1 - 1), deleteL(k2))
  if k1 == k2: return(insertL(i, k1), deleteL(k2 + 1))
  
function XFORML(deleteL(k1), insertL(i, k2)):  
  if k1 < k2: return(deleteL(k1), insertL(i, k2 - 1))
  if k1 > k2: return(deleteL(k1 + 1), insertL(i, k2))
  if k1 == k2: return(deleteL(k1 + 1), insertL(i, k2))
  
function XFORML(deleteL(k1), deleteL(k2)):  
  if k1 < k2: return(deleteL(k1), deleteL(k2 - 1))
  if k1 > k2: return(deleteL(k1 - 1), deleteL(k2))
  if k1 == k2: return(no-op, no-op)    
\end{lstlisting}
\end{figure}

\renewcommand{\ttdefault}{lmtt}

\section{Tree Transformations}\label{sec:main}

In this section we develop a transformation function for $n$-ary tree operations. Therefore we introduce the definition of the transformation point and construct a transformation function that satisfies TP1. The proof that our transformation function satisfies the propterty is stated in the appendix. We maintain a very high level of detail, since the transformation of tree operations requires a precise definition of the transformed access paths.

\begin{sdefinition}{def:TPt}{Transformation Point}
Given two non empty lists $l_1$ and $l_2$ of natural numbers. The \emph{Transformation Point ($\TPt$)} is the index of the first difference of $l_1$ and $l_2$. If $l_1$ is a sublist of $l_2$, the Transformation Point is the index of the last element of $l_1$. If $l_2$ is a sublist of $l_1$, the Transformation Point is the index of the last element of $l_2$.
\end{sdefinition}

If we consider two tree operations, the transformation point marks the point where a transformation may be necessary. We give two short examples of the transformation point:
\[ \TPt\left( \left[ 1,2,3\right] , \left[1,2,4\right]\right) = 2 \hspace*{0.5cm} \TPt\left(\left[1,0\right],\left[1,0,3,2\right]\right) = 1\]

With the transformation point we determine whether two operations are effect dependent or effect independent i.e., if a transformation is necessary or not. We provide a definition for the effect independent tree operations in \defref{def:TConcur}.

\begin{sdefinition}{def:TConcur}{Effect Independence of Tree Operations}
Let $pos_1$ and $pos_2$ be the access paths of the operations $O_1$ and $O_2$ and $tp$ be the transformation point of $pos_1$ and $pos_2$. The operations $O_1$ and $O_2$ are \emph{effect independent tree operations}, denoted by $O_1 \parallel O_2$, iff:
\vspace{-3mm}
\begin{enumerate}
\setlength\itemsep{0em}
  \item $\left(|pos_1| > (tp + 1)\right) \land \left(|pos_2| > (tp + 1)\right)$
  \item $\left(pos_1[tp] > pos_2[tp]\right) \land \left(|pos_1| < |pos_2|\right)$
  \item $\left(pos_1[tp] < pos_2[tp]\right) \land \left(|pos_1| > |pos_2|\right)$
\end{enumerate} 
\end{sdefinition}

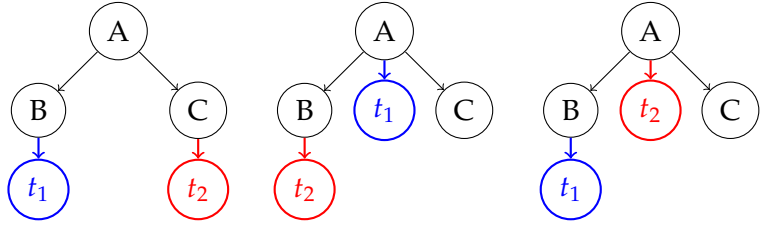
\begin{figure}[t]
  \centering
    \begin{tikzpicture}[auto,bend angle=30,scale=0.7,
      value/.style={draw, circle, minimum height=0.5cm
      }]]
      \node [value](A) at (0,0) {A};
      \node [value](C) at (1.5,-1.5) {C};
      \node [value](B) at (-1.5,-1.5) {B};
      \node [value,color=blue,thick](t1) at (-1.5,-3) {$t_1$};
      \node [value,color=red,thick](t2) at (1.5,-3) {$t_2$};
      
      \path[->] (A) edge node {} (B);
      \path[->] (A) edge node {} (C);
      \path[->,color=blue,thick] (B) edge node {} (t1);
      \path[->,color=red,thick] (C) edge node {} (t2);
      
      \node [value](A2) at (5,0)  {A};
      \node [value](C2) at (6.5,-1.5) {C};
      \node [value](B2) at (3.5,-1.5) {B};
      \node [value,color=blue,thick](t12) at (5,-1.5) {$t_1$};
      \node [value,color=red,thick](t22) at (3.5,-3) {$t_2$};
      
      \path[->] (A2) edge node {} (B2);
      \path[->] (A2) edge node {} (C2);
      \path[->,color=blue,thick] (A2) edge node {} (t12);
      \path[->,color=red,thick] (B2) edge node {} (t22);
      
      \node [value](A3) at (10,0) {A};
      \node [value](C3) at (11.5,-1.5) {C};
      \node [value](B3) at (8.5,-1.5) {B};
      \node [value,color=red,thick](t13) at (10,-1.5) {$t_2$};
      \node [value,color=blue,thick](t23) at (8.5,-3) {$t_1$};
      
      \path[->] (A3) edge node {} (B3);
      \path[->] (A3) edge node {} (C3);
      \path[->,color=red,thick] (A3) edge node {} (t13);
      \path[->,color=blue,thick] (B3) edge node {} (t23);

    \end{tikzpicture}
  \caption{Demonstration of the cases of effect independent tree operations in \defref{def:TConcur}}
  \label{fig:insert-concurrent}
\end{figure}

The three cases of \defref{def:TConcur} are visualized for two $\IT$ operations in \figref{fig:insert-concurrent}. The trees $t_1$ and $t_2$ are the subtrees which should be inserted by the two $\IT$ operations $O_1$ and $O_2$. The effect of the operation $O_1$, that is the insertion of $t_1$, is visualized as a blue circle. The effect of the operation $O_2$ is visualized as a red circle. Note that the Transformation Point in all examples is 0. In the left tree we demonstrate the first case of \defref{def:TConcur}. Both trees $t_1$ and $t_2$ are inserted in nodes which are beyond the transformation point. The trees in the middle and in the right represent the second and third case of \defref{def:TConcur}. In these cases one tree is inserted into a node left to the position where the other tree is inserted. We note that the order of two effect independent operations does not matter.

\begin{sdefinition}{def:UPplus}{$\UP^+$}
The function $\UP^+$ has two input parameters: an access path $pos$ and a number $n$. The result is a modified access path, where the $n$\textsuperscript{th} element of $pos$ is \emph{increased} by $1$. 
\end{sdefinition}

\begin{sdefinition}{def:UPminus}{$\UP^-$}
The function $\UP^-$ has two input parameters: an access path $pos$ and a number $n$. The result is a modified access path, where the $n$\textsuperscript{th} element of $pos$ is \emph{decreased} by $1$. 
\end{sdefinition}

The functions $\UP^+$ and $\UP^-$ are used in the transformation functions to modify an access path of an operation at a specific position. We present the pseudo code for the transformation of an $\IT$ operation against one $\IT$ operation as $\XFORM_T$ in \lstref{lst:insertT_insertT}. Transforming one $\IT$ operation against another one is similar to the transformation of two $\IL$ list operations in $\XFORM_L$. Particularly, in $\XFORM_T$ we perform exact the same transformation at the transformation point as in $\XFORM_L$, only the items of the lists are now subtrees. First we check whether we need a transformation i.e., if both operations are effect independent as defined in \defref{def:TConcur}. Then we check similar to \lstref{lst:XFORML} how the position parameters at the transformation point are related and transform the position parameter at the transformation point as in $\XFORM_L$. If we need to transform one $\IT$ operation against another one with an identical access path, application dependent priorities are used to privilege one operation.

\renewcommand{\ttdefault}{DejaVuSansMono-TLF}

\begin{figure}[t]
\lstset{
basicstyle=\scriptsize\ttfamily, 
frame=top bottom,
numbers=left,
numberstyle=\tiny\ttfamily,
language=Python,
commentstyle=\color{light-gray},
caption={Pseudo code of the transformation of $\IT$ against $\IT$},
label={lst:insertT_insertT}
}
\begin{lstlisting}
function XFORMT(insertT(t1, pos1), insertT(t2, pos2)):
  TP = TPt(pos1, pos2)

  if effectIndependent(pos1, pos2):
    return(insertT(t1, pos1), insertT(t2, pos2))  
    
  if pos1[TP] > pos2[TP]:
    return(insertT(t1, update+(pos1, TP)), insertT(t2, pos2))
    
  if pos1[TP] < pos2[TP]:
    return(insertT(t1, pos1), insertT(t2, update+(pos2,TP)))
    
  if pos1[TP] == pos2[TP]:
    if len(pos1) > len(pos2):
      return(insertT(t1, update+(pos1, TP)), insertT(t2, pos2))

    if len(pos1) < len(pos2):
      return(insertT(t1, pos1), insertT(t2, update+(pos2, TP)))
      
    if pos1 == pos2:
      # use application dependent priorities     
\end{lstlisting}
\end{figure}

\renewcommand{\ttdefault}{lmtt}

We introduce a transformation function for two $\DT$ operations in \lstref{lst:deleteT_deleteT}. The main difference to the previous transformation of $\IT$ operations is that we decrement the position parameters at the transformation point as in $\XFORM_L$. Therefore we use the function $\UP^-$. If the transformation point of both $\DT$ operations is equal we  either delete a subtree from an already deleted subtree or we have two identical position parameters. Both variants are handled with $\operatorname{no-op}$ operations. 

\renewcommand{\ttdefault}{DejaVuSansMono-TLF}

\begin{figure}[t]
\lstset{
basicstyle=\scriptsize\ttfamily, 
frame=top bottom,
numbers=left,
numberstyle=\tiny\ttfamily,
language=Python,
commentstyle=\color{light-gray},
caption={Pseudo code of the transformation of $\DT$ against $\DT$},
label={lst:deleteT_deleteT}
}
\begin{lstlisting}
function XFORMT(deleteT(pos1), deleteT(pos2)):
  TP = TPt(pos1, pos2)
  
  if effectIndependent(pos1, pos2):
    return(deleteT(pos1), deleteT(pos2))
  
  if pos1[TP] > pos2[TP]:
    return(deleteT(update-(pos1, TP)), deleteT(pos2))
    
  if pos1[TP] < pos2[TP]:
    return(deleteT(pos1), deleteT(update-(pos2, TP)))
    
  if pos1[TP] == pos2[TP]:
    if len(pos1) > len(pos2): # delete from a deleted tree
      return(no-op, deleteT(pos2))
    if len(pos1) < len(pos2): # delete from a deleted tree
      return(deleteT(pos1), no-op)
    if pos1 == pos2:
      return(no-op, no-op)
\end{lstlisting}
\end{figure}

\renewcommand{\ttdefault}{lmtt}

After introducing transformation functions for two $\IT$ or two $\DT$ operations, we combine both functions to achieve a transformation function for a transformation of $\IT$ against $\DT$. We state the last transformation function in \lstref{lst:insertT_deleteT}. In the transformation function we modify the access paths exactly as shown in the previous $\XFORM_L$ functions. We observe one special case if the access paths of both operations at the transformation point are identical. If in this case the access path of the $\IT$ operation contains more items than the access path of the $\DT$ operation, one tree is inserted into a deleted tree (corresponding lines \linenumber{14-15}). To solve this conflict, we use a $\noop$ operation as shown in  \lstref{lst:deleteT_deleteT}. 

The the transformation of a $\DT$ operation against an $\IT$ operation can be directly derived from \lstref{lst:insertT_deleteT}. The only difference to he transformation of the $\IT$ against $\DT$ is that the input parameters as well as the return parameters are interchanged. 
  
\renewcommand{\ttdefault}{DejaVuSansMono-TLF}

\begin{figure}[t]
\lstset{
basicstyle=\scriptsize\ttfamily, 
frame=top bottom,
numbers=left,
numberstyle=\tiny\ttfamily,
language=Python,
commentstyle=\color{light-gray},
caption={Pseudo code of the transformation of $\IT$ against $\DT$},
label={lst:insertT_deleteT}
}
\begin{lstlisting}
function XFORMT(insertT(t, pos1), deleteT(pos2)):
  TP = TPt(pos1, pos2)
  
  if effectIndependent(pos1, pos2):
    return(insertT(t, pos1), deleteT(pos2))
  
  if pos1[TP] > pos2[TP]:
    return(insertT(t, update-(pos1, TP)), deleteT(pos2))
    
  if pos1[TP] < pos2[TP]:
    return(insertT(t, pos1), deleteT(update+(pos2, TP)))
    
  if pos1[TP] == pos2[TP]:
    if len(pos1) > len(pos2): # insert into deleted tree
      return(no-op, deleteT(pos2))
    else:
      return(insertT(t, pos1), deleteT(update+(pos2, TP)))    
\end{lstlisting}
\end{figure}

\renewcommand{\ttdefault}{lmtt}

Ultimately we introduced a transformation function for every combination of $\IT$ and $\DT$ operations that fits exactly our needs i.e., the synchronization of edits on generic hierarchical objects. Moreover the stated transformation function is correct with respect to the necessary property (TP1) to guarantee consistent results. We present our formal proof in following subsection.

\subsection{Proofs}

Since our transformation functions is composed of one function for each pair out of $\{\IT, \DT\}$, we proof the TP1 validity for each pair seperately. 

\begin{slemma}{lem:TTP1insertinsert}
The transformation function $\XFORM_T$ satisfies TP1 for the transformation of an $\IT$ against an $\IT$ operation.
\end{slemma}

\begin{lproof}{lem:TTP1insertinsert} Let $(t_1, pos_1,T)$ and $(t_2,pos_2,T)$ be arbitrary but valid input parameters for two $\IT$ operations and let $O_1'$ and $O_2'$ be the result of the transformation of $O_1 = \IT(t_1,pos_1,T)$ against $O_2 = \IT(t_2,pos_2,T)$: \[\XFORM_T\left(O_1, O_2\right) = \left(O_1',O_2'\right)\] 
We show, that $\XFORM_T$ satisfies the TP1 property: \[O_2' \circ O_1 (T) = O_1' \circ O_2 (T)\]
Let $tp = \TPt(pos_1,pos_2)$ be the transformation point and let \[s = T\bigl\llbracket pos_1[< tp]\bigr\rrbracket = (v,L)\] be the subtree at position $pos_1[< tp]$. Note that $pos_1[\le tp]$ and $pos_2[\le tp]$ directing to a position in $L$ and $pos_1[< tp] = pos_2[< tp]$. Let $T'$ be the original tree $T$ without the subtree $s$. For each if-statement in $\XFORM_T$ we consider the proof separately and highlight the corresponding lines in \lstref{lst:insertT_insertT} for each case.

\begin{case}
$O_1 \parallel O_2$ (corresponding line \linenumber{2})

If $O_1$ and $O_2$ are effect independent, the order of the execution does not influence the result. According to $\XFORM_T$ both operations are left untouched during the transformation. We show that $O_2 \circ O_1 (T) = O_1 \circ O_2 (T)$.

We have already given some intuition for effect independent cases in \figref{fig:insert-concurrent}. Next we prove that the execution order is independent for such operations. 

Let $T_1 = \IT(t_1,pos_1,T)$ be the tree, where the subtree \[T\llbracket pos_1[<|pos_1| - 1] \rrbracket = (v_1,L_1)\] is replaced by $(v_1,\IL(t_1,pos_1[|pos_1| - 1],L_1))$. Note that the subtree $(v_1,L_1)$ is the subtree where $t_1$ should be inserted into. 

Let $T_2 = \IT(t_2,pos_2,T)$ be the subtree where \[T\llbracket pos_2[<|pos_2| - 1] \rrbracket = (v_2,L_2)\] is replaced by $(v_2,\IL(t_2,pos_2[|pos_2| - 1],L_2))$.  We restate the proof goal for this case to:
\[\IT(t_2,pos_2,T_1) = \IT(t_1,pos_1,T_2)\] 

Consider the tree $\IT(t_2,pos_2,T_1)$. We expect a new tree i.e., the result, where the subtree \[T_1\llbracket pos_2[<|pos_2| - 1] \rrbracket = (v_2',L_2')\] is replaced by $(v_2',\IL(t_2,pos_2[|pos_2| - 1],L_2'))$. Because of the conditions in \defref{def:TConcur} we note that for any non-empty sublist $pos_2'$ of $pos_2$ we have neither inserted nor deleted a tree to a position left of $pos_2'[|pos_2'| - 1]$. In contrast to situations where we have two $\IL$ operations on the same list, we have never performed a ``shift'' during the insertion of $t_1$. Hence $(v_2',L_2') = (v_2,L_2)$. We obtain a tree where $(v_1,L_1)$ is replaced by \[(v_1,\IL(t_1,pos_1[|pos_1| - 1],L_1))\] and $(v_2,L_2)$ is replaced by \[(v_2,\IL(t_2,pos_2[|pos_2| - 1],L_2))\]

From this point we can analogously show that the obtained tree is identical to $\IT(t_1,pos_1,T_2)$. Thus $O_2 \circ O_1 (T) = O_1 \circ O_2 (T)$.
\end{case}

\begin{case}
$pos_1[tp] > pos_2[tp]$ (corresponding lines \linenumber{7-8})
  
We have the transformed operations:
\begin{eqnarray*}
O_1' & =& \IT(t_1,\UP^+(pos_1,tp),O_1)\\
O_2' &=& \IT(t_2,pos_2,O_2)
\end{eqnarray*}
Since $O_1$ and $O_2$ are not effect dependent, we know that $pos_2[tp]$ refers to the last element of $pos_2$.
Hence we restate our proof goal: 
\begin{eqnarray*}
& & \IT(t_2,pos_2,\IT(t_1,pos_1,T)) \\
&= & \IT(t_1,\UP^+(pos_1,tp),\IT(t_2,pos_2,T))
\end{eqnarray*}
We divide the proof goal into two sub cases since $\IT$ is recursively defined in \defref{def:insertT}. Both cases are visualized in \figref{fig:pos1pos2}. In this example the transformation point is 0 and from $pos_1[tp] > pos_2[tp]$ we know that $t_1$ is inserted at a position right of $t_2$. We show the goals directly by using the presented lemmata and definitions.

\begin{figure}[t]
\centering
\begin{tikzpicture}[auto,bend angle=30,node distance=2cm,scale=.7]
\node [draw, circle, minimum height=0.7cm](A) at (0,0)  {A};
\node [draw, circle, minimum height=0.7cm](C) at (0.8,-2) {C};
\node [draw, circle, minimum height=0.7cm](B) at (-0.8,-2) {B};
\node [draw, circle, minimum height=0.7cm,color=blue,thick](t1) at (2.4,-2) {$t_1$};
\node [draw, circle, minimum height=0.7cm,color=red,thick](t2) at (-2.4,-2) {$t_2$};

\node at (0,1.1) {$|pos_1| = |pos_2|$};

\path[->] (A) edge node {} (B);
\path[->] (A) edge node {} (C);
\path[->,color=blue,thick] (A) edge node {} (t1);
\path[->,color=red,thick] (A) edge node {} (t2);

\node at (6,1.1) {$|pos_1| > |pos_2|$};        

\node [draw, circle, minimum height=0.7cm](A3) at (6,0) {A};
\node [draw, circle, minimum height=0.7cm](C3) at (7.6,-1.5) {C};
\node [draw, circle, minimum height=0.7cm](B3) at (6,-1.5) {B};
\node [draw, circle, minimum height=0.7cm,color=red,thick](t13) at (4.4,-1.5) {$t_2$};
\node [draw, circle, minimum height=0.7cm,color=blue,thick](t23) at (7.6,-3) {$t_1$};

\path[->] (A3) edge node {} (B3);
\path[->] (A3) edge node {} (C3);
\path[->,color=red,thick] (A3) edge node {} (t13);
\path[->,color=blue,thick] (C3) edge node {} (t23);

\end{tikzpicture}
\caption{Example situations for case 2 in the proof of \lemref{lem:TTP1insertinsert}}
\label{fig:pos1pos2}
\end{figure}

\begin{subcase}
$|pos_1| = |pos_2|$

In this case we insert the subtrees $t_1$ and $t_2$ into the subtree $s$. 
{\scriptsize 
\begin{eqnarray*}
&& \IT(t_2,pos_2,\IT(t_1,pos_1,T)) \\[-1mm]
&\stackrel{\text{\tiny L\ref{lem:insertT}}}{=} & \IT(t_2,pos_2,\IT(\IT(t_1,pos_1[\ge tp],s),pos_1[< tp],T')) \\[-1mm]
&\stackrel{\text{\tiny D\ref{def:insertT}}}{=} & \IT(t_2,pos_2,\IT((v,\IL(t_1,pos_1[tp],L)),pos_1[< tp],T'))\\[-1mm]    
&\stackrel{\text{\tiny L\ref{lem:insertT}}}{=} & \IT(\IT(t_2,pos_2[\ge tp],(v,\IL(t_1,pos_1[tp],L))),pos_1[< tp],T')\\[-1mm]
&\stackrel{\text{\tiny D\ref{def:insertT}}}{=} & \IT((v,\IL(t_2,pos_2[tp],\IL(t_1,pos_1[tp],L))),pos_1[< tp],T')\\[-1mm]
&\stackrel{\text{\tiny TP1}}{=} & \IT((v,\IL(t_1,pos_1[tp] + 1,\IL(t_2,pos_2[tp],L))),pos_1[< tp],T')\\[-1mm]
&\stackrel{\text{\tiny D\ref{def:insertT}}}{=} & \IT(\IT(t_1,\UP^+(pos_1,tp)[\ge tp],( v ,\IL(t_2,pos_2[tp],L))),\\[-1mm]
&& \hspace{0.75cm} pos_2[< tp],T')\\[-1mm]
&\stackrel{\text{\tiny L\ref{lem:insertT}}}{=} & \IT(t_1,\UP^+(pos_1,tp),\IT(( v , \IL(t_2,pos_2[tp],L)),\\[-1mm]
&&\hspace{0.75cm} pos_2[< tp],T'))\\[-1mm]
&\stackrel{\text{\tiny D\ref{def:insertT}}}{=} & \IT(t_1,\UP^+(pos_1,tp),\IT(\IT(t_2,pos_2[\ge tp],s),\\[-1mm]
&&\hspace{0.75cm} pos_2[< tp],T'))\\[-1mm]
&\stackrel{\text{\tiny L\ref{lem:insertT}}}{=} & \IT(t_1,\UP^+(pos_1,tp),\IT(t_2,pos_2,T))
\end{eqnarray*}}
In the 4\textsuperscript{th} line of the proof we referring to the TP1 validity of $\XFORM_L$.
\end{subcase}

\begin{subcase}
$|pos_1| > |pos_2|$

In this case we insert the subtree $t_2$ into the subtree $s$ and replace the tree at position $pos_1[\le tp]$ with a new subtree which contains $t_1$. Let $L'$ be the list of subtrees $L$ without the subtree at position $pos_1[\le tp]$ and let $s_{O_1}$ be the subtree at position $pos_1[\le tp]$ after executing $O_1$ to the tree.
{\scriptsize
\begin{eqnarray*}
L' & = & \DL(pos_1[tp],L) \\ [1.2mm]
s_{O_1} & = & \IT(t_1,pos_1,T)\bigl\llbracket pos_1[\le tp]\bigr\rrbracket   \\
  & = & \IT(t_1,pos_1[>tp],L \bigl[ pos_1\left[tp\right] \bigr] )\\
  & = & \IT(t_1,pos_1[>tp],\IL(t_2,pos_2[tp],L) \bigl[ pos_1\left[tp\right] + 1 \bigr] )
\end{eqnarray*}}
We show our proof goal directly by using the presented lemmata and definitions.
{\scriptsize 
\begin{eqnarray*}
&  & \IT(t_2,pos_2,\IT(t_1,pos_1,T)) \\[-1mm]
&\stackrel{\text{\tiny L\ref{lem:insertT}}}{=} & \IT(t_2,pos_2,\IT(\IT(t_1,pos_1[\ge tp],s),pos_1[< tp],T')) \\[-1mm]
&\stackrel{\text{\tiny D\ref{def:insertT}}}{=} & \IT(t_2,pos_2,\IT((v,\IL(s_{O_1},pos_1[tp],L')),pos_1[< tp],T'))\\[-1mm]    
&\stackrel{\text{\tiny L\ref{lem:insertT}}}{=} & \IT(\IT(t_2,pos_2[\ge tp],(v,\IL(s_{O_1},pos_1[tp],L'))),pos_1[< tp],T')\\[-1mm]
&\stackrel{\text{\tiny D\ref{def:insertT}}}{=} & \IT((v,\IL(t_2,pos_2[tp],\IL(s_{O_1},pos_1[tp],L'))),pos_1[< tp],T')\\[-1mm]
&\stackrel{\text{\tiny TP1}}{=} & \IT((v,\IL(s_{O_1},pos_1[tp] + 1,\\[-1mm]
&& \hspace{0.75cm}\IL(t_2,pos_2[tp],L'))),pos_1[< tp],T')\\[-1mm]
&\stackrel{\text{\tiny TP1}}{=} & \IT((v,\IL(s_{O_1},pos_1[tp] + 1,\\[-1mm]
&& \hspace{0.75cm} \DL(pos_1[tp] + 1,\IL(t_2,pos_2[tp],L)))), pos_1[< tp],T')\\[-1mm]
&\stackrel{\text{\tiny D\ref{def:insertT}}}{=} & \IT(\IT(t_1,\UP^+(pos_1,tp)[\ge tp],\\[-1mm]
&& \hspace{0.75cm}( v , \IL(t_2,pos_2[tp],L))),pos_2[< tp],T')\\[-1mm]
&\stackrel{\text{\tiny L\ref{lem:insertT}}}{=} & \IT(t_1,\UP^+(pos_1,tp),\\[-1mm]
&& \hspace{0.75cm}\IT(( v , \IL(t_2,pos_2[tp],L)),pos_2[< tp],T'))\\[-1mm]
&\stackrel{\text{\tiny D\ref{def:insertT}}}{=} & \IT(t_1,\UP^+(pos_1,tp),\\[-1mm]
&& \hspace{0.75cm}\IT(\IT(t_2,pos_2[\ge tp],s),pos_2[< tp],T'))\\[-1mm]
&\stackrel{\text{\tiny L\ref{lem:insertT}}}{=} & \IT(t_1,\UP^+(pos_1,tp),\IT(t_2,pos_2,T))
\end{eqnarray*}}
\end{subcase}
\vspace*{-12pt}
\end{case}

\begin{case} 
$pos_1[tp] < pos_2[tp]$ (corresponding lines \linenumber{10-11})

This case is analog to the previous case. 
\end{case}

\begin{case}
$pos_1[tp] = pos_2[tp]$ (corresponding lines \linenumber{13-21})

The scenarios of this case are demonstrated in \figref{fig:case4TTP1}. We analyze the three cases separately.
\begin{figure*}[t]
  \centering
    \begin{tikzpicture}[auto,bend angle=30,node distance=2cm,scale=.7]
      \node [draw, circle, minimum height=.7cm](A) at (0,0) {A};
      \node [draw, circle, minimum height=.7cm](C) at (1.5,-1.5) {C};
      \node [draw, circle, minimum height=.7cm](B) at (-1.5,-1.5) {B};
      \node [draw, circle, minimum height=.7cm,color=blue,thick](t1) at (1.5,-3) {$t_1$};
      \node [draw, circle, minimum height=.7cm,color=red,thick](t2) at (0,-1.5) {$t_2$};
    
      \node at (0,1.1) {$|pos_1| > |pos_2|$};
      \path[->] (A) edge node {} (B);
      \path[->] (A) edge node {} (C);
      \path[->,color=blue,thick] (C) edge node {} (t1);
      \path[->,color=red,thick] (A) edge node {} (t2);
    
      \node [draw, circle, minimum height=.7cm](A2) at (5,0)  {A};
      \node [draw, circle, minimum height=.7cm](C2) at (6.5,-1.5) {C};
      \node [draw, circle, minimum height=.7cm](B2) at (3.5,-1.5) {B};
      \node [draw, circle, minimum height=.7cm,color=blue,thick](t12) at (5,-1.5) {$t_1$};
      \node [draw, circle, minimum height=.7cm,color=red,thick](t22) at (6.5,-3) {$t_2$};      
    
      \node at (5,1.1) {$|pos_1| < |pos_2|$};
      
      \path[->] (A2) edge node {} (B2);
      \path[->] (A2) edge node {} (C2);
      \path[->,color=blue,thick] (A2) edge node {} (t12);
      \path[->,color=red,thick] (C2) edge node {} (t22);
    
      \node [draw, circle, minimum height=.7cm](A3) at (10,0) {A};
      \node [draw, circle, minimum height=.7cm](C3) at (11.5,-2) {C};
      \node [draw, circle, minimum height=.7cm](B3) at (8.5,-2) {B};
      \node [draw, circle, minimum height=.7cm,color=red,thick](t13) at (9.95,-2) {$t_2$};
      \node [draw, circle, minimum height=.7cm,color=blue,thick](t23) at (10.05,-2) {$t_1$};
    
      \node at (10,1.1) {$|pos_1| = |pos_2|$};
      \path[->] (A3) edge node {} (B3);
      \path[->] (A3) edge node {} (C3);
      \path[->,color=red,thick] (A3) edge node {} (t13);
      \path[->,color=blue,thick] (A3) edge node {} (t23);
    \end{tikzpicture}
  \caption{Example situations for case 4 in the proof of \lemref{lem:TTP1insertinsert}}
  \label{fig:case4TTP1}
\end{figure*}

\begin{figure*}[ht]
    \centering
      \begin{tikzpicture}[auto,bend angle=30,node distance=2cm,scale=.7]
        \node [draw, circle, minimum height=.7cm](A) at (0,0) {A};
        \node [draw, circle, minimum height=.7cm,color=red,thick,dotted](C) at (1.5,-1.5) {C};
        \node [draw, circle, minimum height=.7cm](B) at (-1.5,-1.5) {B};
        \node [draw, circle, minimum height=.7cm,color=blue,thick,dotted](t1) at (1.5,-3) {D};
      
        \node at (0,1.1) {$|pos_1| > |pos_2|$};
        \path[->] (A) edge node {} (B);
        \path[->,color=red,thick,dotted] (A) edge node {} (C);
        \path[->,color=blue,thick,dotted] (C) edge node {} (t1);
      
        \node [draw, circle, minimum height=.7cm](A2) at (5,0)  {A};
        \node [draw, circle, minimum height=.7cm,color=blue,thick,dotted](C2) at (6.5,-1.5) {C};
        \node [draw, circle, minimum height=.7cm](B2) at (3.5,-1.5) {B};
        \node [draw, circle, minimum height=.7cm,color=red,thick,dotted](t22) at (6.5,-3) {D};      
      
        \node at (5,1.1) {$|pos_1| < |pos_2|$};
        
        \path[->] (A2) edge node {} (B2);
        \path[->,color=blue,thick,dotted] (A2) edge node {} (C2);
        \path[->,color=red,thick,dotted] (C2) edge node {} (t22);
      
        \node [draw, circle, minimum height=.7cm](A3) at (10,0) {A};
        \node [draw, circle, minimum height=.7cm](C3) at (11.5,-2) {D};
        \node [draw, circle, minimum height=.7cm](B3) at (8.5,-2) {B};
        \node [draw, circle, minimum height=.7cm,color=red,thick,dotted](t13) at (10,-2) {C};
      
        \node at (10,1.1) {$|pos_1| = |pos_2|$};
        \path[->] (A3) edge node {} (B3);
        \path[->] (A3) edge node {} (C3);
        \path[->,color=red,thick,dotted] (A3) edge node {} (t13);
      \end{tikzpicture}
    \caption{Examples for case 4 in the proof of \lemref{lem:TTP1deletedelete}}
    \label{fig:case3DD}
  \end{figure*}

\begin{subcase}
$|pos_1| > |pos_2|$ (corresponding lines \linenumber{14-15})

In this case we insert the tree $t_2$ directly left to the subtree $s_{O_1}$ (the subtree at position $pos_1[\le tp]$ after applying $O_1$ to $T$). This case is analog to case 2.
\end{subcase}

\begin{subcase}
$|pos_1| < |pos_2| $ (corresponding lines \linenumber{17-18})

In this case we insert the tree $t_1$ directly left to the subtree $s_{O_2}$ (the subtree at position $pos_2[\le tp]$ after applying $O_2$ to $T$). This case is analog to case 3.
\end{subcase}

\begin{subcase}
$|pos_1| = |pos_2|$ (corresponding lines \linenumber{20-21})

In this case we need to insert the trees $t_1$ and $t_2$ directly to the same position. We necessarily need to decide which operation should be preferred. We can use application dependent priorities to handle the case as in case 2 or case~3.
\end{subcase}
\end{case}\qed
\end{lproof}

For the following proofs we reduce the level of detail since the substantial parts are demonstrated in the proof of \lemref{lem:TTP1insertinsert}.

\begin{slemma}{lem:TTP1deletedelete}
The transformation function $\XFORM_T$ satisfies TP1 for the transformation of a $\DT$ against a $\DT$ operation.
\end{slemma}

\begin{lproof}{lem:TTP1deletedelete}
\setcounter{case}{0}
We have two delete operations $O_1$ and $O_2$ with $O_1=\DT(pos_1,T)$ and $O_2=\DT(pos_2,T)$ and the operations from the transformation function $\XFORM_T(O_1,O_2) = (O_1',O_2')$ and divide the proof in one case per if-statement in \lstref{lst:deleteT_deleteT}.
\begin{case}
$O_1 \parallel O_2$ (corresponding line \linenumber{2})

The proof of this case is analog to the proof of the corresponding case in \lemref{lem:TTP1insertinsert}, except the $\IL$ operations are exchanged by $\DL$ operations.
\end{case}

\begin{case}
$pos_1[tp] > pos_2[tp]$ (corresponding lines \linenumber{7-8})

After applying the transformation function we can restate the precise proof goal to:
\begin{eqnarray*}
& & \DT(pos_2,\DT(pos_1,T)) \\ [-1mm]
&=& \DT(\UP^-(pos_1,tp),\DT(pos_2,T))
\end{eqnarray*}

As in the previous lemma, we get two sub cases for from the definition of $\DT$. The two sub cases are illustrated in \figref{fig:case1DD}. Either we delete two trees from the same list or we delete a tree from a tree right to the deleted subtree. The effect of $O_1$ is visualized by a dotted blue circle and $O_2$ is visualized by a dotted red circle.

We use \lemref{lem:deleteT}, \defref{def:deleteT} and the Transformation Property 1 of $\XFORM_L$ to show the proof goal directly for each case.

\begin{figure}[t]
    \centering
      \begin{tikzpicture}[auto,bend angle=30,node distance=2cm,scale=.7]
        \node [draw, circle, minimum height=0.7cm](A) at (0,0)  {A};
        \node [draw, circle, minimum height=0.7cm](C) at (0,-2) {C};
        \node [draw, circle, minimum height=0.7cm,color=red,thick,dotted](B) at (-1.6,-2) {B};
        \node [draw, circle, minimum height=0.7cm,color=blue,thick,dotted](t1) at (1.6,-2) {D};
        
        \node at (0,1.1) { $|pos_1| = |pos_2|$};
      
        \path[->,color=red,thick,dotted] (A) edge node {} (B);
        \path[->] (A) edge node {} (C);
        \path[->,color=blue,thick,dotted] (A) edge node {} (t1);
      
        \node at (6,1.1) {$|pos_1| > |pos_2|$};        
        
        \node [draw, circle, minimum height=0.7cm](A3) at (6,0) {A};
        \node [draw, circle, minimum height=0.7cm](C3) at (7.6,-1.5) {D};
        \node [draw, circle, minimum height=0.7cm](B3) at (6,-1.5) {C};
        \node [draw, circle, minimum height=0.7cm,color=red,thick,dotted](t13) at (4.4,-1.5) {B};
        \node [draw, circle, minimum height=0.7cm,color=blue,thick,dotted](t23) at (7.6,-3) {E};
      
        \path[->] (A3) edge node {} (B3);
        \path[->] (A3) edge node {} (C3);
        \path[->,color=red,thick,dotted] (A3) edge node {} (t13);
        \path[->,color=blue,thick,dotted] (C3) edge node {} (t23);
      \end{tikzpicture}
    \caption{Examples for case 2 in the proof of \lemref{lem:TTP1deletedelete}}
    \label{fig:case1DD}
  \end{figure}

\end{case}

\begin{case} 
$pos_1[tp] < pos_2[tp]$ (corresponding lines \linenumber{10-11})

This case is analog to the previous case. Only the operations are interchanged.
\end{case}

\begin{case}
$pos_1[tp] = pos_2[tp]$ (corresponding lines \linenumber{13-19})

There are three scenarios for this case which are demonstrated in \figref{fig:case3DD}. In the first scenario the operation $O_1$ deletes a tree from a tree that is deleted by $O_2$. According to the transformation function the $O_1$ is eliminated and will be transformed to $\operatorname{no-op}$. The second scenario is identical only the operations are interchanged. In the third scenario we have identical position parameters for $O_1$ and $O_2$. Hence one tree is deleted by both operations. According to the transformation function we achieve a consistent result if both operations are transformed to $\operatorname{no-op}$. The proof of this case is analog to the proof of the corresponding case in \lemref{lem:TTP1insertinsert}. \qed

\end{case}
\end{lproof}

Since the transformation function for the transformation of a $\DT$ operation against an $\IT$ operation and vice versa is derived from \lstref{lst:insertT_insertT} and \lstref{lst:deleteT_deleteT}, we omit the proof of TP1 for this case since the substantial parts are already shown in the previous lemmata. Ultimately we declare the final theorem.

\begin{stheorem}{thm:TTP1}
The transformation function $\XFORM_T$ satisfies TP1.
\end{stheorem}

\begin{tproof}{thm:TTP1} The transformation of $\IT$ against $\IT$ satisfies TP1 as shown in \lemref{lem:TTP1insertinsert}. From \lemref{lem:TTP1deletedelete} we know that the transformation of $\DT$ against $\DT$ satisfies TP1 as well. The proof of TP1 in the transformation of $\IT$ against $\DT$ and $\DT$ against $\IT$ is analog to the proof of \lemref{lem:TTP1insertinsert} and \ref{lem:TTP1deletedelete}. Hence all cases of $\XFORM_T$ are satisfying TP1. \qed
\end{tproof}

\section{Conclusion}
With the introduced transformation functions for ordered $n$-ary trees we presented a fruitful way to use operational transformation to synchronize replicas of such structures in an optimistic and comfortable way. Since we have focused on such generic data types, the range of applications that could implement our ideas is wide spread. The transformation function for trees can easily be adopted to other hierarchal architectures like XML documents, which are commonly used in web services and other applications to store and exchange objects. 

We have analyzed the correctness of the transformation functions for lists and $n$-ary trees. All analyzed transformation functions satisfy TP1 which is essential for a successful synchronization with operational transformation. In addition we implemented a proof of TP1 for the transformation function for lists in the interactive theorem prover Isabelle/HOL so we can ensure that our results are correct up to the correctness of Isabelle/HOL.
 
As future work, the proof of TP1 for the transformation function for trees can be implemented in Isabelle/HOL to ensure the correctness of our proof up to the correctness of Isabelle/HOL. In addition our transformation functions can be integrated in an OT programming framework so that applications can benefit from our approach. 

One more complex future work is the enhancement of the transformation functions so that the more complex Transformation Property 2 (TP2) is satisfied. If TP2 is satisfied, more algorithms and less restrictive algorithms can be used to synchronize replicas of trees. 

\bibliographystyle{abbrv}

\begin{thebibliography}{10}

\bibitem{Davis:2002:GOT:587078.587088}
A.~H. Davis, C.~Sun, and J.~Lu.
\newblock Generalizing operational transformation to the standard general
  markup language.
\newblock In {\em Computer Supported Cooperative Work}, CSCW '02, pages 58--67,
  2002.

\bibitem{Ellis:1989:CCG:66926.66963}
C.~A. Ellis and S.~J. Gibbs.
\newblock Concurrency control in groupware systems.
\newblock {\em SIGMOD Rec.}, 18(2):399--407, 1989.

\bibitem{Imine:2003:PCT:1241889.1241904}
A.~Imine, P.~Molli, G.~Oster, and M.~Rusinowitch.
\newblock Proving correctness of transformation functions in real-time
  groupware.
\newblock In {\em European Conference on Computer Supported Cooperative Work},
  ECSCW '03, pages 277--293, Norwell, MA, USA, 2003.

\bibitem{Imine2006167}
A.~Imine, M.~Rusinowitch, G.~Oster, and P.~Molli.
\newblock Formal design and verification of operational transformation
  algorithms for copies convergence.
\newblock {\em Theoretical Computer Science}, 351(2):167 -- 183, 2006.

\bibitem{JuHeACM}
T.~Jungnickel and T.~Herb.
\newblock Simultaneous editing of json objects via operational transformation.
\newblock In {\em ACM Symposium of Applied Computing}, SAC '16, 2016.

\bibitem{Nichols:1995:HLW:215585.215706}
D.~A. Nichols, P.~Curtis, M.~Dixon, and J.~Lamping.
\newblock High-latency, low-bandwidth windowing in the jupiter collaboration
  system.
\newblock In {\em Symposium on User Interface and Software Technology}, UIST
  '95, pages 111--120, 1995.

\bibitem{OsterICEIS07}
G.~Oster, H.~Skaf-Molli, P.~Molli, and H.~Naja-Jazzar.
\newblock {Supporting Collaborative Writing of {XML} Documents}.
\newblock In {\em Enterprise Information Systems: Software Agents and Internet
  Computing}, pages 335--342, 2007.

\bibitem{Ressel:1996:ITA:240080.240305}
M.~Ressel, D.~Nitsche-Ruhland, and R.~Gunzenh\"{a}user.
\newblock An integrating, transformation-oriented approach to concurrency
  control and undo in group editors.
\newblock In {\em Computer Supported Cooperative Work}, CSCW '96, pages
  288--297, 1996.

\bibitem{4530334}
H.~Skaf-Molli, P.~Molli, C.~Rahhal, and H.~Naja-Jazzar.
\newblock Collaborative writing of xml documents.
\newblock In {\em Information and Communication Technologies: From Theory to
  Applications}, ICTTA '08, pages 1--6, 2008.

\bibitem{Sun:2006:TAS:1188816.1188821}
C.~Sun, S.~Xia, D.~Sun, D.~Chen, H.~Shen, and W.~Cai.
\newblock Transparent adaptation of single-user applications for multi-user
  real-time collaboration.
\newblock {\em ACM Trans. Comput.-Hum. Interact.}, 13(4):531--582, 2006.

\bibitem{Sun:2014:ESP:2531602.2531630}
C.~Sun, Y.~Xu, and A.~Agustina.
\newblock Exhaustive search of puzzles in operational transformation.
\newblock In {\em Proceedings of the 17th ACM Conference on Computer Supported
  Cooperative Work \& Social Computing}, CSCW '14, pages 519--529, New York,
  NY, USA, 2014. ACM.

\end{thebibliography}

\end{document}